\documentclass[prc,showpacs,showkeys,superscriptaddress,nofootinbib,floatfix,twocolumn,groupedaddress]{revtex4}
\usepackage{color}
\usepackage{amsfonts}
\usepackage{amsbsy}
\usepackage{mathrsfs}
\usepackage{graphicx}
\def\lsim{\mathrel{\rlap{
\lower4pt\hbox{\hskip-3pt$\sim$}}
    \raise1pt\hbox{$<$}}}     
\def\gsim{\mathrel{\rlap{
\lower4pt\hbox{\hskip-3pt$\sim$}}
    \raise1pt\hbox{$>$}}}     
\def\scr#1{\mbox{\scriptsize #1}}
\begin{document}
\title{Collective Flow in Heavy-Ion Collisions from AGS to SPS}
\author{V.N.~Russkikh}\thanks{e-mail: russ@ru.net}
\affiliation{Gesellschaft
 f\"ur Schwerionenforschung mbH, Planckstr. 1,
D-64291 Darmstadt, Germany}
\affiliation{Kurchatov Institute, Kurchatov
sq. 1, Moscow 123182, Russia}
\author{Yu.B.~Ivanov}\thanks{e-mail: Y.Ivanov@gsi.de}
\affiliation{Gesellschaft
 f\"ur Schwerionenforschung mbH, Planckstr. 1,
D-64291 Darmstadt, Germany}
\affiliation{Kurchatov Institute, Kurchatov
sq. 1, Moscow 123182, Russia}
\begin{abstract}
Collective transverse flow in heavy-ion collisions at incident energies  
$E_{\scr{lab}}\simeq$ (1--160)$A$ GeV is analyzed within the model of 3-fluid 
dynamics (3FD). Simulations are performed with purely
hadronic equation of state (EoS). 
At the AGS energies the flow turns out to be sensitive 
to the stopping power of nuclear 
matter rather than only to the stiffness of the EoS. 
When the stopping power is fixed to reproduce other observables, 
the flow data favor
more and more soft EoS with the incident energy rise, which can be
associated with ``a transition from hadronic to string
matter'' reported in 
the Hadron-String-Dynamics (HSD) model.  
Problems, which are met in simultaneous reproduction of 
directed and elliptic flows within the 3FD, suggest that the
transverse flow is very sensitive to the character of the
transverse-momentum {\it nonequilibrium} at the initial stage of
collision. 
Arguments in favor of "early-stage" nature of 
the flow observable are put forward. This suggests that the flow 
(especially the directed one) is 
determined by early-stage evolution of the collision rather than 
freeze-out stage. 

\end{abstract}
%

\maketitle

\section{Introduction}

The interest to nucleus-nucleus collisions at incident energies
$E_{\scr{lab}}\simeq$ (10--40)$A$ GeV has been recently revived,
since the highest baryon densities 
\cite{Friman98,CB,3FD} and highest relative strangeness
\cite{Gazdzicki99,BCOR02} at 
moderate temperatures are expected in this energy range.
The onset of deconfinement is also expected in this domain. In
particular, the energy-scan SPS program \cite{SPS} is dedicated to the
search for the onset of deconfinement in heavy-ion collisions. 
Moreover, 
a critical end point \cite{Asakawa} 
of the QCD phase diagram
may be accessible in these
reactions \cite{Fodor01,Stephanov99}. 
The above expectations motivated the project of the new
accelerator facility FAIR at GSI \cite{SIS200}, the heavy-ion program
of which is precisely dedicated to studying dense baryonic matter with
the emphasis on the onset of deconfinement and the critical end point. 
The future SPS \cite{NA49-future} and RHIC \cite{RHIC-future} programs
are also devoted to the same problems. 

In Refs. \cite{3FD,3FDm} we have introduced 3-fluid 
dynamical (3FD) model which is suitable for simulating heavy-ion
collisions in the range from AGS to SPS energies, which overlaps the
energy range of the future FAIR. We have started
simulations \cite{3FD}  with the purely hadronic equation of state (EoS)
\cite{gasEOS} in order to see how good the available data from AGS and
SPS can be understood without involving the concept of deconfinement. 
This EoS should serve as a reference point for simulations involving
more sophisticated EoS's including the phase transition.

With this simple hadronic EoS we succeeded to reasonably
reproduce a great 
body of experimental data in the incident energy range
$E_{\scr{lab}}\simeq$ (1--160)$A$ GeV. The list includes 
 rapidity distributions,  transverse-mass spectra, and 
multiplicities of various hadrons. 
However, we also found
out certain problems. In particular, we 
failed to describe the transverse
flow  at $E_{\scr{lab}}\geq 40A$ GeV.

One of the main  conclusions of Ref. \cite{3FD} is 
that in order to conclude on the
relevance of a particular EoS the whole available set of data should
be analyzed in a wide incident energy range 
with the same fixed parameters of the model. Indeed almost for any
particular piece of data the parameters of the model can be tuned in
such a way that the model reasonably reproduces the data. At the same
time the reproduction of the whole set of available data {\em with the
  same model parameters} is far from being always possible. In
Ref. \cite{3FD} we performed  still fragmentary
analysis of data due to constraints to frames of a single
paper. Therefore, in the present paper we would like to continue our
analysis of a particular subset of the data concerning collective flow
in nuclear collisions, based on the set of model parameters fixed in
\cite{3FD}. Accordingly to our preliminary results 
\cite{3FD}, this subset is the most problematic for the purely hadronic
EoS we use. 

The collective transverse flow is formed at the early (compression)
stage of the 
collisions and hence reveals the early pressure gradients in the
evolving nuclear matter. The harder EoS is, the stronger pressure is
developed. Thus, the flow reflects 
stiffness of the nuclear EoS. The collective flow observable has been
extensively exploited to obtain information on the EoS 
\cite{cass90,Stoecker,gut89,Gale2,Zhang,Russ95,EOS95,her96,cha97,dani98,%
sorge97,rei97,E877,Dani,mar94,Rischke,bas97,SahuNS,bli99,Brac00a,%
hom98,nara97,Frank1,Fuchs,sahu98,Sahu00,Larionov,INNTS,Sahu02,%
DLL02,IOSN05,H01,LPX99,BBSG99,Roehrich99,Zabrodin01,SBBSZ04}. 
In particular, it was expected that the transition to the
quark-gluon phase results in significant reduction of the directed
flow \cite{Hung95}
(so called ``softest point'' effect), since the pressure in the
quark-gluon phase is lower than that in the hadronic
phase. Calculations within conventional (1-fluid) hydrodynamics
\cite{Rischke} indicated almost disappearance of the flow at the 
onset of the first-order phase transition. However, the nuclear finite
stopping power essentially weakens this effect
\cite{Brac00a,INNTS}. In the present paper we are going to 
discuss the 
transverse flow in terms of stiffness of the nuclear EoS
and the stopping power achieved in the collision. 

At present there are quite systematic collective-flow data, both on
directed and elliptic transverse flows, at AGS 
\cite{FOPI05,E895-02,E877-px,E877-v2,E895-99,E895-00,E895-02b} 
and SPS \cite{NA49-98-v1,NA49-99,NA49-03-v1,CERES:INPC01}
energies. We confine our 
present discussion to the flow of non-strange probes, i.e. nucleons and
pions. Recent analysis of this set of data was done in
Ref. \cite{IOSN05} with emphasis on momentum dependence of
nuclear forces. This was done within 
the JAM model \cite{jam} combined with a
covariant prescription of mean fields (RQMD/S) \cite{RQMD_S} with
the emphasis on the momentum dependence of nuclear forces. 
We would like to analyze the same data but within the hydrodynamic
approach.

\section{3FD Model}

A direct way to address thermodynamic properties of the matter
produced in these reactions consists in application of
hydrodynamic simulations to nuclear collisions. However, finite
nuclear stopping power, revealing itself at high incident
energies, makes the collision dynamics non-equilibrium and
prevents us from application of conventional hydrodynamics
especially at the initial stage of the reaction. Since the
resulting  non-equilibrium is quite strong, introduction of
viscosity and thermal conductivity does not help to overcome this
difficulty, because by definition they are suitable for weak
non-equilibrium. A possible way out is taking advantage of a
multi-fluid approximation to heavy-ion collisions.

Unlike the conventional hydrodynamics, where local instantaneous
stopping of projectile and target matter is assumed, a specific
feature of the dynamic 3-fluid description is a finite stopping
power resulting in a counter-streaming regime of leading
baryon-rich matter.  Experimental rapidity distributions in
nucleus--nucleus  collisions support this counter-streaming
behavior, which can be observed for incident energies  between few
and 200$A$ GeV. The basic idea of a 3-fluid approximation to
heavy-ion collisions \cite{3FD,3FDm} is that at each space-time
point $x=(t,{\bf x})$ the generally nonequilibrium 
distribution function of baryon-rich
matter, can be represented as a sum of two
distinct contributions, 
$f_{\scr{bar.}}(x,p)=f_{\scr p}(x,p)+f_{\scr t}(x,p)$,
initially associated with constituent nucleons of the projectile
(p) and target (t) nuclei. In addition, newly produced particles,
populating the mid-rapidity region, are associated with a fireball
(f) fluid described by the  distribution function $f_{\scr f}(x,p)$. 
It is assumed that constituents within each
distribution are  locally equilibrated, both thermodynamically and
chemically. This assumption, justifying the term ``fluids'', 
relies on the
fact that  intra-fluid collisions are much more efficient in
driving a system to equilibrium than  inter-fluid interactions.
Therefore, the 3-fluid approximation is a minimal way to 
simulate the finite stopping power at high incident energies.
Note that both the baryon-rich and fireball fluids may
consist of any type of hadrons  and/or partons (quarks and
gluons), rather than only nucleons and pions.

Our 3FD model \cite{3FD,3FDm} is a
straightforward extension of the 2-fluid model with radiation of
direct pions \cite{MRS88,gsi94,gsi91} and (2+1)-fluid model
\cite{Kat93,Brac97}. We extend the above models in such a
way that the created baryon-free fluid (which we call a
``fireball'' fluid, following the Frankfurt group) is treated
on equal footing with the baryon-rich ones. 
We allow a certain formation time for the fireball fluid, during
which the matter of the fluid propagates without interactions. 
We assume that the fireball matter gets quickly thermalized
after its formation. The latter approximation is an enforced one, since
we deal with the hydrodynamics rather than with kinetics.

\begin{figure}[htb]
\includegraphics[width=8cm]{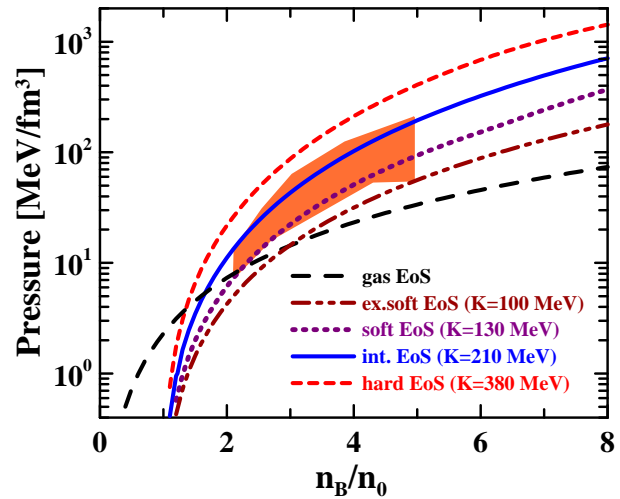}
$\;$\vspace*{-2mm} 
\caption{ (Color online) Baryon-density dependence of the pressure at
  zero temperature for hard ($K=$ 380 MeV), intermediate ($K=$
  210 MeV), soft ($K=$ 130 MeV), extrasoft ($K=$ 100 MeV), 
and gas EoS's. Shaded   region is the constraint 
  derived from experimental data in Ref. \cite{DLL02}. } 
\label{fig1}
\end{figure}

The main unknowns of the present simulations can be briefly
summarized as follows: an EoS and ``cross sections''.
Our goal is to find an EoS which in the best way reproduces
{\em the largest body of  available observables}. The ``cross sections'' are
equally important. They determine friction forces between fluids
and hence the nuclear stopping power. 
In principle, friction
forces are EoS dependent, because medium modifications, providing a
nontrivial EoS, also modify cross sections, and should be
externally supplied together with the EoS. However, 
at present we
have at our disposal only a rough estimate of the friction forces
\cite{Sat90}. 
Therefore, we have to fit the
friction forces to the stopping power observed in proton rapidity
distributions.

We have started our simulations \cite{3FD}
with a simple, purely hadronic EoS \cite{gasEOS} which involves only
a density dependent mean field providing saturation of cold
nuclear matter at normal nuclear density $n_0=$ 0.15 fm$^{-3}$ and
with the proper binding energy -16 MeV, cf. Appendix 
\ref{Hadronic EoS}. This EoS is a natural 
reference point for any other more elaborate EoS.

Fig. \ref{fig1} displays the density dependence of 
the pressure of four versions of this hadronic EoS at zero
temperature: hard EoS  with incompressibility $K=$ 380 MeV,
intermediate EoS  with $K=$ 210 MeV, soft EoS with $K=$ 130 MeV, and 
extrasoft EoS with $K=$ 100 MeV. 
The pressure of the nucleon gas is also presented, indicating the
softest EoS within our parametrization. 
As seen from Fig. \ref{fig1}, the intermediate EoS, with which all the
calculations of Ref.  \cite{3FD} were done,  and the soft EoS, 
are within the constraint given by Danielewicz {\it et al.}
\cite{DLL02} based on 
the analysis of flow of  nuclear matter  at the AGS
incident energies.

\begin{figure}[htb]
\includegraphics[width=8cm]{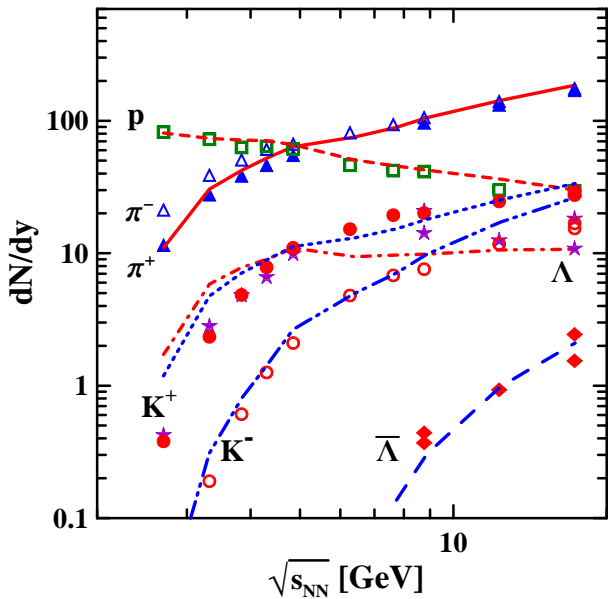}
$\;$\vspace*{-2mm} 
\caption{(Color online) Incident energy dependence of midrapidity
  yield of various 
  hadrons produced in central nucleus--nucleus collisions. At the AGS
  energies the 3FD calculations with intermediate EoS ($K=$ 210 MeV) 
were done for Au+Au collisions at impact
  parameter $b=$ 2 fm, at SPS energies --- for Pb+Pb at $b=$ 2.5 fm.
The compilation of experimental data is taken from Ref. \cite{Andronic}.} 
\label{fig2}
\end{figure}
\begin{figure}[htb]
\includegraphics[width=7cm]{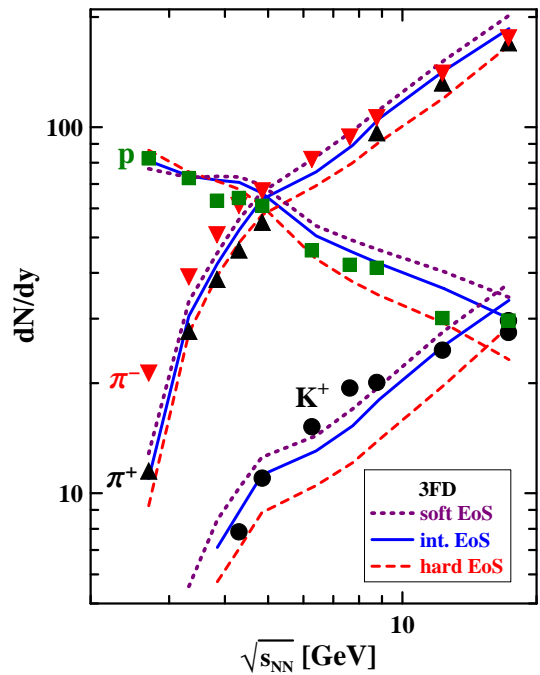}
$\;$\vspace*{-2mm} 
\caption{(Color online) 
The same as in Fig. \ref{fig2} but zoomed and for three EoS's displayed in
Fig. \ref{fig1}. } 
\label{fig2a}
\end{figure}

Strictly speaking,
the constraint region of Ref. \cite{DLL02} is valid the class of
EoS's considered in \cite{DLL02}, since the flow essentially depends
also on the temperature dependence of the pressure, which is not
displayed in Fig. \ref{fig1}. Moreover, since the incompressibility
characterizes the EoS only in the vicinity of the bound state,
different EoS's with the same incompressibility may differ at higher
densities even at zero temperature. 
This is the reason why the incompressibility of our intermediate EoS 
($K=$ 210 MeV) differs from that of the 
intermediate EoS of Ref. \cite{DLL02} ($K=$ 300 MeV), while these two
EoS's are very similar at densities higher than $2n_0$. 
The same concerns soft EoS's: $K=$ 130 MeV in our case, and $K=$ 210
MeV in Ref. \cite{DLL02}. Of course, values $K=$ 100 and 130 MeV are
unrealistically small from point of view of low-energy nuclear
physics. However, in simulating heavy-ion collisions we 
predominantly deal with
densities $n \gsim 2n_0$. Therefore, we should estimate the degree of
reasonableness of a EoS precisely in this density region,
considering the incompressibility only as label for the EoS.

A comprehensive
analysis of the nuclear EoS \cite{Klahn06}, based on astrophysical data, 
indicated that the preferable region of
Fig. \ref{fig1} should be slightly shifted to higher pressures. 
Thus, the hard version of our EoS also falls into the region of EoS's 
reasonable from the astrophysical point of view. At the same time, our
soft and extrasoft EoS's fall out of this astrophysically acceptable region.

The 3FD model with the intermediate EoS  turned out to be able
to reasonably 
reproduce a great body of experimental data \cite{3FD} in a wide
energy range from AGS to SPS. 
Fig. \ref{fig2} illustrates (and extends) the results of Ref. \cite{3FD}. 
The experimental data were compiled in Ref. \cite{Andronic} based on
reported experimental results 
\cite{E877:piKp,E895:piKp,E917:piKp,NA49:piKp,NA44:piKp,%
NA57:piKp,E895-Lambda,E917-Lambda,NA49-Lambda}. 
These data slightly differ in degree of centrality, at which they were
taken. We however performed our calculations for Au+Au collisions at
fixed impact 
  parameter $b=$ 2 fm for the AGS energies, and for Pb+Pb at $b=$ 2.5
  fm for the SPS energies. Slight irregularity of calculated curves
  between $\sqrt{s_{NN}}=4.85$ GeV and $\sqrt{s_{NN}}=6.27$ GeV (i.e. between
$E_{\scr{lab}}=10A$ GeV and $E_{\scr{lab}}=20A$ GeV) is related to
transition to different colliding nuclei at different impact
  parameter. 
Poor description of $K^+$ and $\Lambda$ yields at low incident
energies is not surprising, since our EoS is based on the grand
canonical ensemble. Really surprising is the fact that $K^-$ and
$\bar{\Lambda}$ yields are still well reproduced at low energies.

In the present paper we also do calculations with hard, soft and
estrasoft EoS's
(cf. Fig. \ref{fig1}) in order to study effects of the EoS stiffness. 
Changes, introduced by these hard and soft EoS's,  
are presented in Fig. \ref{fig2a}. As seen, the intermediate EoS is
indeed preferable in overall reproduction of the data. 

To study effect of nuclear stopping power on the transverse flow, 
we also do calculations within the
conventional 1-fluid dynamics (1FD), for which we have a separate
code.

\section{Directed Flow}
\label{Directed Flow}

The conventional transverse-momentum flow of an $a$ species is defined 
as \cite{DO85}
\begin{eqnarray}
\langle p_x^{(a)}\rangle (y)= 
\frac{\displaystyle \int d^2 p_T \ p_x  E \ dN_a/d^3p }%
{\displaystyle \int d^2 p_T E \ dN_a/d^3p}, 
\label{eq-px}
\end{eqnarray}
where $p_x$ is the transverse momentum of a particle in the reaction
plane, $E \ dN_a/d^3p$ is the momentum distribution of $a$-hadrons,
which takes due account of feed-down from resonance decays, 
and integration runs over the transverse momentum $p_T$.

\begin{figure}[ht]
\includegraphics[width=8cm]{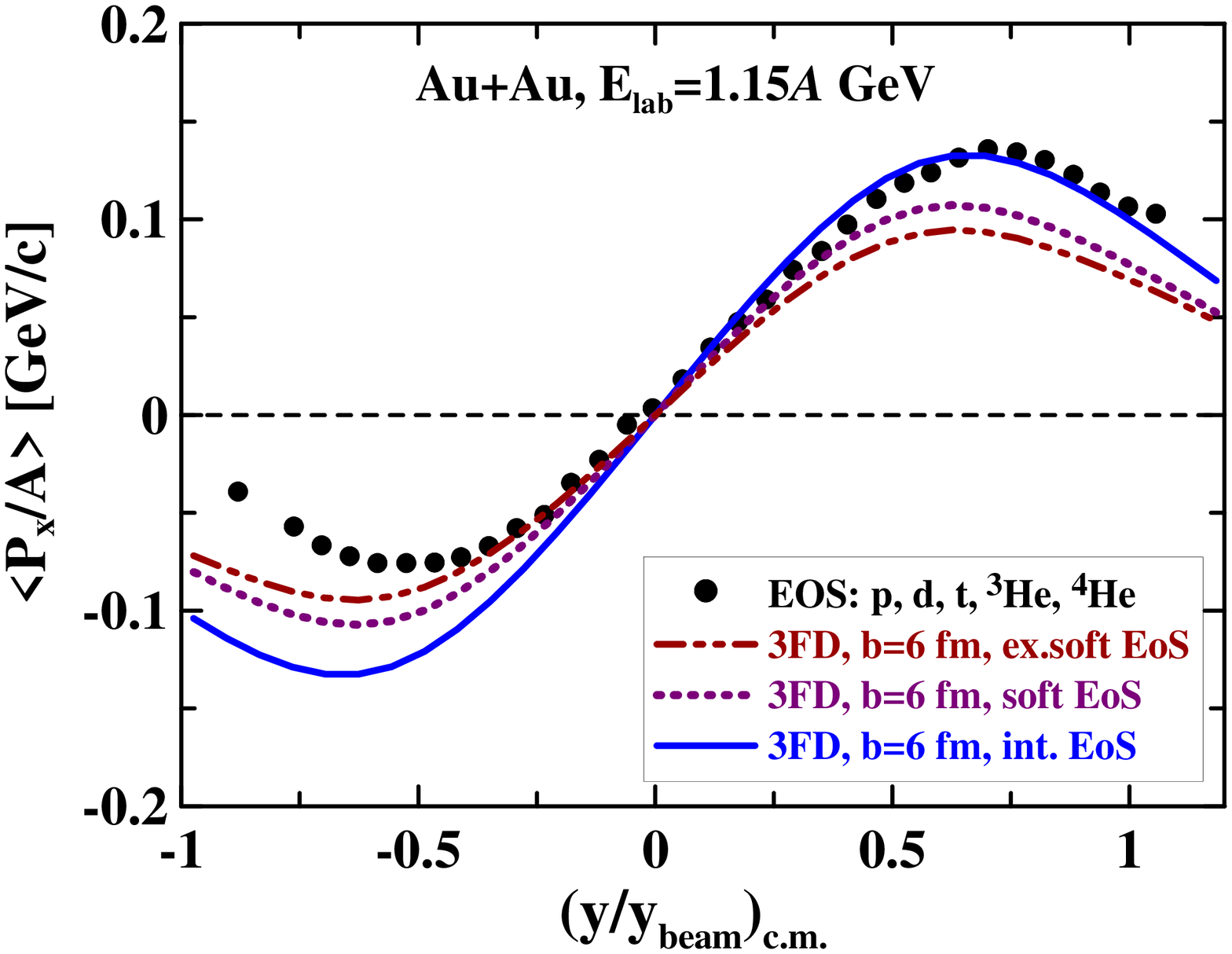}
\caption{(Color online) 
Directed flow of nucleons as a
  function of rapidity 
for mid-central Au+Au collisions at $E_{\scr{lab}}=$ 1.15$A$
GeV. The 3FD results at $b=$ 6 fm are presented for three EoS's. 
The data are from Ref. \cite{EOS95}.
} 
\label{fig3}
\end{figure}
\begin{figure*}[htb]
\includegraphics[width=11.0cm]{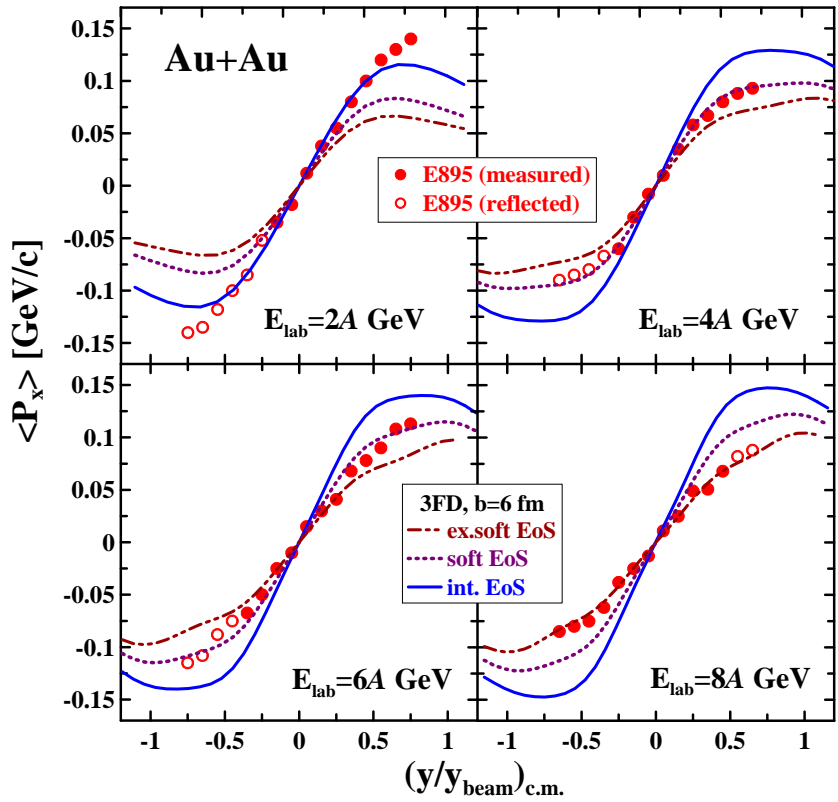}
\caption{(Color online) 
Directed flow of protons as a function of rapidity 
for mid-central Au+Au collisions at $E_{\scr{lab}}= 2A, 4A, 6A$ and $8A$
GeV. The 3FD results at $b=$ 6 fm are presented for three EoS's. 
The data are from Ref. \cite{E895-00}.
} 
\label{fig4}
\end{figure*}
\begin{figure}[htb]
\includegraphics[width=8.cm]{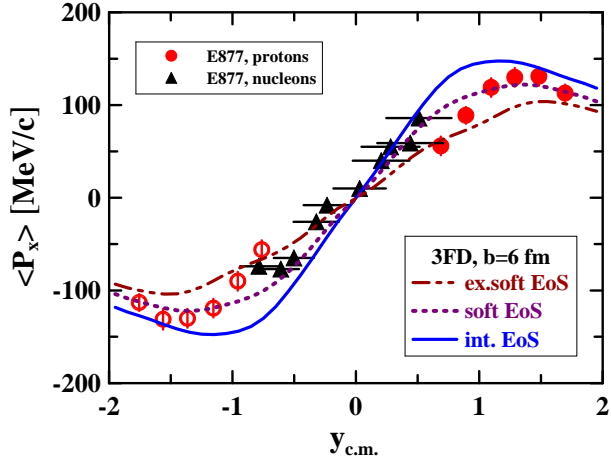}
\caption{(Color online) Directed flow of nucleons and protons as a
  function of rapidity 
for mid-central Au+Au collisions at $E_{\scr{lab}}=$ 10.5$A$
GeV. The 3FD calculations with three EoS's
are presented for impact parameter $b=$ 6 fm. 
Data are from Ref. \cite{E877-px}.
Full symbols correspond to measured data, open ones are those reflected 
with respect to the midrapidity point. } \label{fig5}
\end{figure}

In Figs. \ref{fig3}--\ref{fig5} the comparison with Bevalac and AGS
data on the directed flow of protons \cite{EOS95,E895-00,E877-px} and
``nucleons'' \cite{EOS95,E877-px} is presented. Under ``nucleon data''
we mean the combined flow of identified protons and light nuclear
fragments \cite{EOS95,E877-px}. 
In fact, we compute $\langle p_x\rangle$ of so-called
primordial nucleons, which later may coalesce, forming light
fragments. 
In view of this, the comparison with the nucleon 
data is preferable for our model, since it
avoids additional assumptions of coalescence. 
In the case of $E_{\scr{lab}}=$ 10.5$A$ GeV (cf. Fig. \ref{fig5}),
agreement with nucleon data indeed 
seems to be better than that with identified-proton data. 
Since experimentally the sideward flow is observed for mid-central
collisions with an average impact parameter of $b=$ 6 fm, we restrict
our computation to this single impact parameter here.

It is seen that the observed directed flow prefers 
more and more soft EoS with the incident energy rise. While at 
$E_{\scr{lab}}\lsim 2A$ GeV the intermediate EoS is preferable,
even the soft EoS looks too hard at $E_{\scr{lab}}\gsim 8A$ GeV. 
At $E_{\scr{lab}}= 8A$ GeV the extrasoft EoS is certainly preferable,
while at $E_{\scr{lab}}= 10.5A$ GeV it is difficult to choose between
soft and extrasoft EoS's. 
However, in
Refs. \cite{Sahu02,IOSN05,BBSG99}, a rather weak sensitivity
of the flow to the stiffness of the EoS was reported.
An apparent reason for this disagreement is that in above models the
EoS becomes effectively softer with
incident energy rise. At higher incident energies, 
this softening occurs because of gradual (governed by form-factors) 
\cite{Sahu02} or abrupt \cite{BBSG99} switching off mean-field
interactions, or because more and more
initial baryon--baryon collisions at the early collision stage 
end up in strings which are not
affected by the nuclear mean field \cite{IOSN05}. 
Since the mean-field switching off is accompanied by dominating string
excitations also in models \cite{Sahu02,BBSG99}, it is possible to
summarize this mechanism as ``a transition from hadronic to string
matter'', following Ref. \cite{Sahu00}.  This
early-stage ``string matter'' results in effective softening of the
EoS. 

\begin{figure}[bht]
\includegraphics[width=6.5cm]{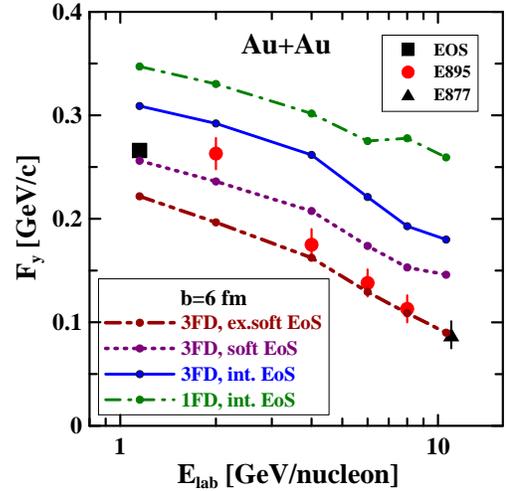}
\caption{(Color online) Proton flow magnitude as a
  function of beam energy 
for mid-central Au+Au collisions. 
The 3FD calculations are presented for 
and $b=$ 6 fm.  Compilation of the data is from Ref. \cite{E895-00}.
} \label{fig6}
\end{figure}
\begin{figure*}[bht]
\includegraphics[width=12.cm]{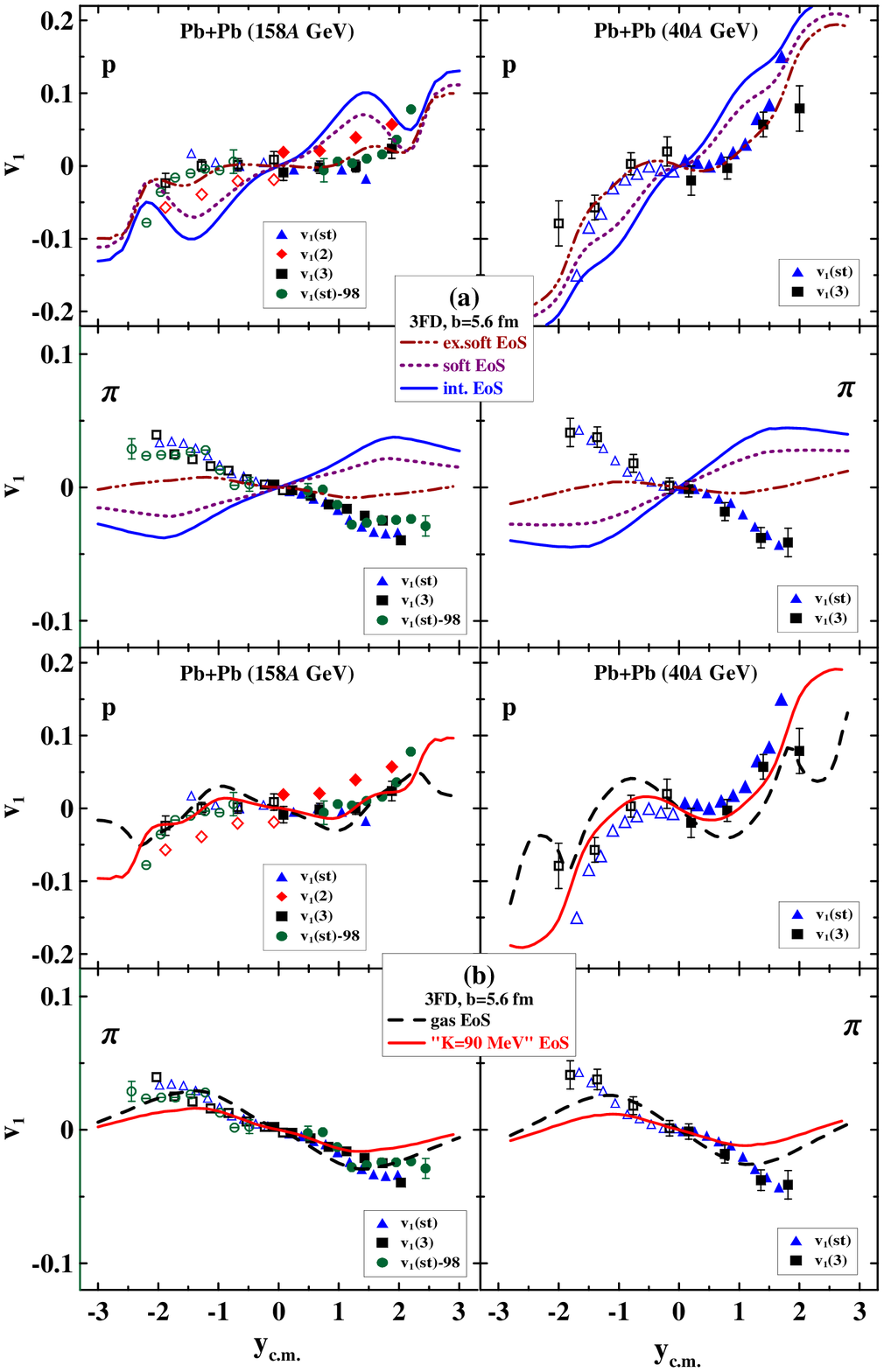}
\caption{(Color online) Directed
flow of protons (upper panels) and charged pions (lower panels) in 
mid-central Pb+Pb collisions at $E_{\scr{lab}}=$ 158$A$ (left panels) and 
40$A$ (right panels)  GeV as a function of rapidity. In the upper block
of panels (a) the 3FD calculations at $b=$ 5.6 fm are presented 
for intermediate, soft and extrasoft EoS's; at the lower block (b) ---
the gas EoS and EoS with $K=90$ MeV. 
Experimental data  Ref. \cite{NA49-03-v1} obtained by two different
methods are displayed: by the standard method ($v(st)$) 
and by the method of $n$-particle correlations ($v(n)$). 
Full symbols correspond to
measured data, while open symbols are those reflected  
with respect to the midrapidity. Updated data of the NA49
Collaboration \cite{NA49-98-v1} ($v(st)-98$) with acceptance 
0.05 $<p_T<$ 0.35 GeV/c for pions  and
0.6 $<p_T<$ 2.0  Gev/c for protons, are also shown.}  
\label{fig7}
\end{figure*}

The flow magnitude is conventionally defined as the slope of directed
flow at midrapidity
\begin{eqnarray}
F_y^{(a)} = 
\left(\frac{d\langle p_x^{(a)}\rangle}{dy}\right)_{y=y_{cm}}. 
\label{Fy}
\end{eqnarray}
The excitation function of the proton flow magnitude is shown in
Fig. \ref{fig6}. This figure summarizes observations
made above. In fact, the point at $E_{\scr{lab}} = 1.15A$ GeV
is somewhat misleading. While the differential directed flow is well
reproduced at this energy, cf. Fig. \ref{fig3}, the slope is not, 
because of asymmetry of the data with respect to midrapidity. 
Note that for the symmetric system like Au+Au 
the flow should be antisymmetric with respect to midrapidity. 
Additionally Fig. \ref{fig6} demonstrates that the nuclear stopping
power also essentially affects the directed flow. The 
conventional 1-fluid simulations (infinite stopping power) result in
essentially stronger flow than it is observed.

At high energies the azimuthal asymmetry is usually characterized by
the first coefficients of Fourier expansion of the
azimuthal-angle dependence of the
 single-particle distribution function, 
i.e. by the directed flow $v_1=\langle \cos \phi\rangle$ \cite{Voloshin96}: 
\begin{eqnarray}
 \label{eq-v1}
v_1^{(a)} (y)&=& 
\frac{\displaystyle \int d^2 p_T  \left(p_x/p_T\right)
E \ dN_a/d^3p }%
{\displaystyle \int d^2 p_T 
E \ dN_a/d^3p }
. 
\end{eqnarray}
Examples of $v_1$ flow for protons and pions at 
$E_{\scr{lab}}=$ 40$A$ and 158$A$ GeV are presented in Fig. 
\ref{fig7} for mid-central collisions. 
The 3FD simulations are performed at fixed impact parameter as  $b=$ 5.6 fm. 
Several sets of data, which noticeably
differ from each other, are shown in these figures. 
The "standard" method \cite{DO85,Voloshin96,PV98} for evaluating the flow
 coefficients requires an event-by-event estimate of the reaction
 plane with which outgoing hadrons
  correlate. However, this method does not discriminate other sources of
 correlations, like those due to global momentum conservation,
 resonance decays, etc.  A new method \cite{BDO01} of
$n$-particle correlations allows to get
 rid of these non-flow correlations in extracting $v_1$ and $v_2$ from
 genuine azimuthal correlations.

At these energies the proton $v_1$ flow, calculated with intermediate
and soft EoS's,  significantly differs from the data. 
The experimental data favor  
extrasoft EoS, which became already preferable at $E_{\scr{lab}}=8A$
GeV, cf. Fig. \ref{fig4}. 
Thus the EoS softening with the incident energy rise stops near
the top AGS energy, at higher energies the same extrasoft EoS
remains preferable. 
This fits the observation in Ref. \cite{Sahu00} that the
``transition from hadronic to string matter'', i.e. early-stage
softening of the EoS, practically saturates at the top AGS energy. 
Apparently, the early-stage softening of the EoS at SPS energies 
is the reason of the success of microscopic RQMD~\cite{LPX99}, 
UrQMD~\cite{BBSG99},  
JAM \cite{IOSN05} and HSD \cite{SBBSZ04} models.  
These models reveal quite soft dynamics at the early stage, 
since mean fields turn out to be essentially reduced in this energy
range.

The lower block of panels (b) in Fig. \ref{fig7} demonstrates what
happens if we further soften the EoS. Slight softening of the
extrasoft EoS, i.e. taking $K=$ 90 MeV, results in better reproduction
of pion $v_1$ however on the expense of proton flow. The softest EoS,
i.e. the gas EoS (cf. Fig. \ref{fig1}), already again contradicts the
data, since it produces a wiggle in proton $v_1$, which is not
experimentally observed.

The calculated pion directed flow closely follows the pattern of 
the proton one, while the pion $v_1$ data reveal anticorrelation with 
proton flow. 
In fact, our pion directed flow correlates with proton one even at
lower, AGS energies, instead of anticorrelating. 
A probable reason for this poor reproduction of pion $v_1$ consists in
disregarding the fact that a part of frozen-out
particles is ``shadowed'' by still hydrodynamically evolving matter.  
This mechanism of proton-pion anticorrelation was discussed in 
Refs. \cite{Bass93,Li94}. 
In terms of hydrodynamics, 
this shadowing means that frozen-out particles cannot 
freely propagate through the region still occupied by the hydrodynamically 
evolved matter but rather get reabsorbed into the hydrodynamic phase.
%
%
Apparently, the baryon directed flow is less affected by this 
shadowing. The reason is that 
the baryon directed flow reveals the collective flow of matter, 
since this matter 
is mainly built of baryons as the most abundant and heavy component of
the system. This collective flow is mainly formed at the early stage
of the reaction. Baryon rescatterings within this earlier-formed
collective flow at later stages do not 
essentially alter the collective transverse momentum of the matter. 
At the same time, the pions are much stronger
affected by this   
shadowing, since they are screened by the predominantly baryonic matter, 
where pions may rescatter or even be absorbed. This can
drastically  change the pion $v_1$.

\section{Elliptic Flow}
\label{Elliptic Flow}

The elliptic flow, $v_2=\langle \cos 2 \phi\rangle$ \cite{Voloshin96}, is 
the second coefficient of Fourier expansion of the
azimuthal-angle dependence of the
 single-particle distribution function 
\begin{eqnarray}
 \label{eq-v2}
v_2^{(a)} (y)&=& 
\frac{\displaystyle \int d^2 p_T \ \left[(p^2_x- p^2_y)/p^2_T\right]
E \ dN_a/d^3p }%
{\displaystyle \int d^2 p_T 
E \ dN_a/d^3p }
. 
\end{eqnarray}

\begin{figure}[thb]
\includegraphics[width=7cm]{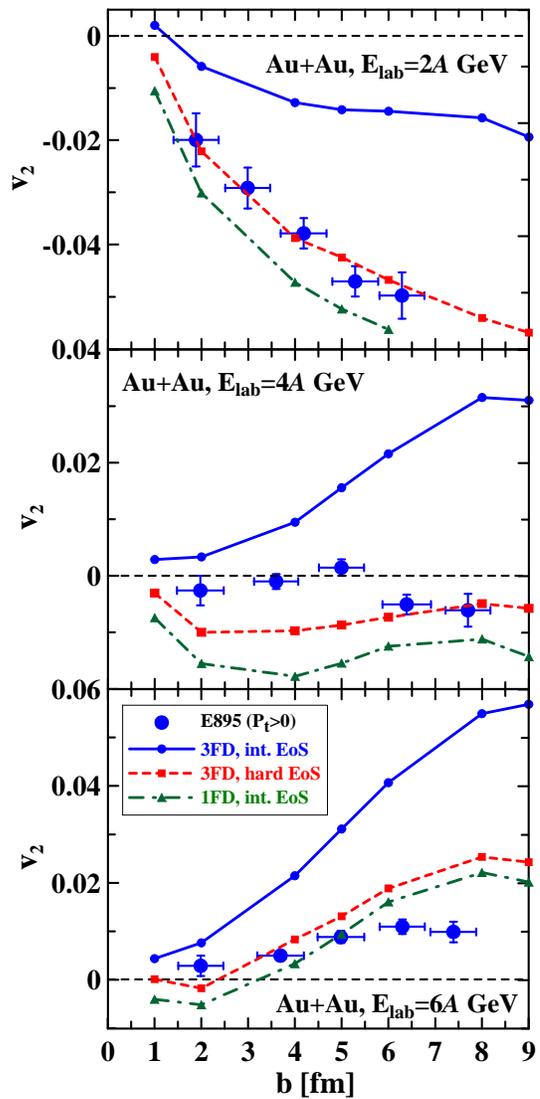}
\caption{(Color online) 
Midrapidity elliptic flow as a function of impact parameter
in Au+Au collisions at $E_{\scr{lab}}=2A,4A$ and 6$A$ GeV. 
Experimental data for $p_T>0$ are from Ref. \cite{E895-02}. 
}
\label{fig8}
\end{figure}
\begin{figure}[thb]
\includegraphics[width=6.5cm]{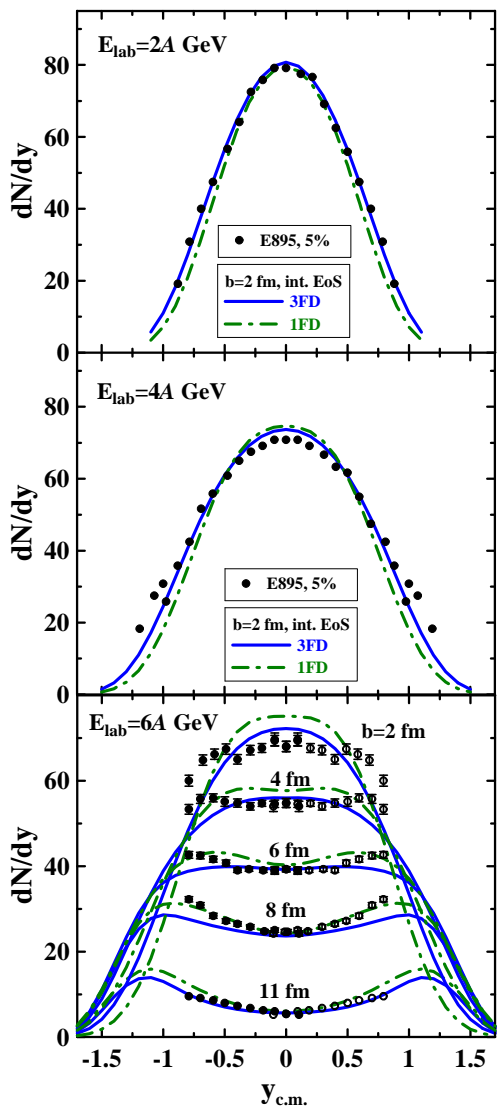}
\caption{(Color online) Proton rapidity spectra at AGS energies for 
various impact parameters. Solid lines correspond to 3FD 
calculations with intermediate EoS. For
comparison the 1FD result with the same EoS is shown by the dotted-dashed line. 
Experimental points are taken from 
\cite{E895:piKp} at 2$A$ and 4$A$ GeV,  
and \cite{E917:piKp} at 6$A$ GeV. The percentage indicates 
the fraction of the total reaction cross section, 
corresponding to experimentally selected events.} 
\label{fig9}
\end{figure}

Impact parameter dependence of the proton
elliptic flow for Au+Au collisions at energies $E_{\scr{lab}}=2A,4A$
and 6$A$ GeV without any cut on the transverse momentum ($p_T>0$) is
presented in Fig. \ref{fig8}. 
We see that the soft and intermediate EoS's, which are the best in
reproducing the directed flow, are completely off the
experimental data. At the same time, the hard EoS reasonably reproduces
$v_2$. This observation is similar to that done in
Refs. \cite{dani98,DLL02,E895-02}. In Refs. \cite{Sahu02,IOSN05,SBBSZ04}
it was found that a proper momentum dependence 
in the nuclear mean field is of prime importance for the 
simultaneous reproduction of the directed and elliptic flows.

The fact that this momentum dependence is vitally important apparently
again implies that the initial stage of nuclear collision is essentially
nonequilibrium and that the flow is sensitive to this early
nonequilibrium in transverse-momentum distribution.  
Indeed, even if a momentum dependent forces are
present in the model underlying a EoS, in the EoS this momentum dependence is
integrated out with the equilibrium momentum distribution. Therefore,
the in- and out-of-reaction-plane transverse pressures, which are
responsible for 
$v_1$ and $v_2$ flows, are tightly interrelated. This prevents the
1FD model from simultaneous reproduction of
$v_1$ and $v_2$. In the 3FD model these two components of pressure are
less tightly interrelated but still do not possess the same degree of
freedom as in genuine nonequilibrium, since a superposition of three
equilibrium distributions certainly confines the class of possible
nonequilibrium momentum configurations. The 3-fluid
approximation was designed to primarily simulate the
transverse-longitudinal anisotropy of momentum distribution.
It was not specially tuned to describe 
anisotropy within the transverse direction, which is of prime
importance for $v_2$.
Apparently, the genuine
momentum nonequilibrium at the initial stage together with proper
momentum dependence in the nuclear mean field is 
required for simultaneous reproduction of the directed and elliptic
flows.

In view of above said, our intermediate EoS, which
is the best in overall reproduction of a large body of other data,
can be considered 
as a kind of compromise allowing to simultaneous
reproduce orders of magnitude of the directed and elliptic flows
within the assumptions of the 3FD model.

\begin{figure}[htb]
\includegraphics[width=8cm]{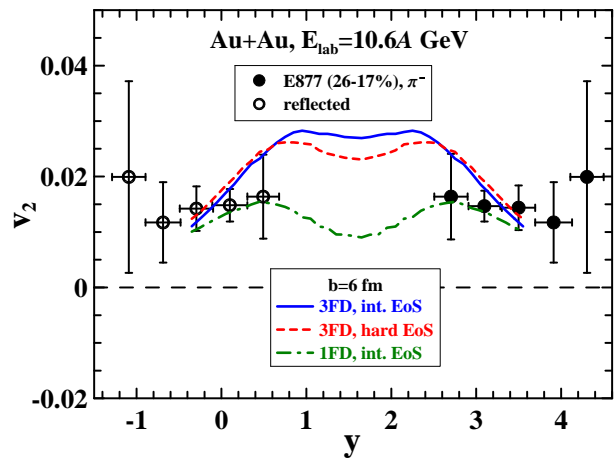}
\caption{(Color online) 
Pion elliptic flow as a function of rapidity
in Au+Au collisions at $E_{\scr{lab}}=11.5A$ GeV. 
Experimental data for $p_T>0$ are from Ref. \cite{E877-v2}. 
}
\label{fig10}
\end{figure}
\begin{figure*}[ht]
\includegraphics[width=12.5cm]{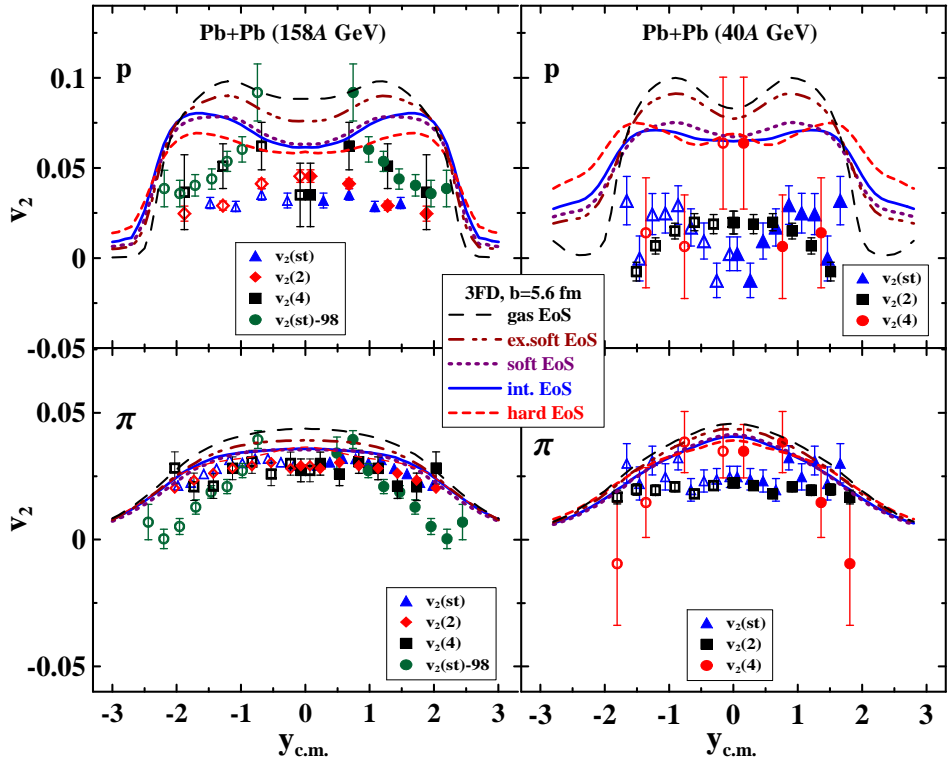}
\caption{(Color online) 
The same as in Fig. \ref{fig7} but for the elliptic flow.}
\label{fig7a}
\end{figure*}
\begin{figure}[thb]
\includegraphics[width=6.5cm]{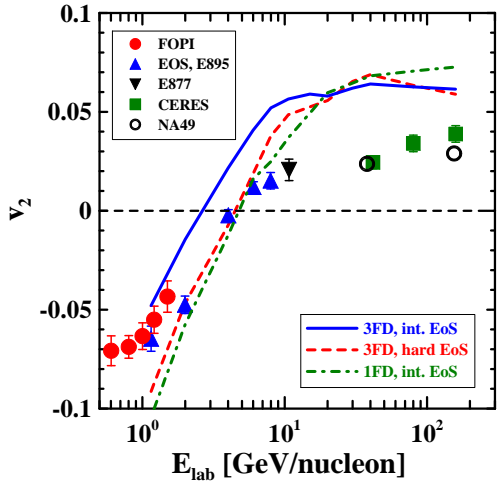}
\caption{(Color online) 
Proton elliptic flow at midrapidity as a function of incident energy
in mid-central  
Au+Au (at AGS energies) and Pb+Pb (at SPS energies) collisions. 
3FD calculations with intermediate and hard EoS's, as well as 1FD
calculations are presented. Compilation of experimental data is from 
Ref. \cite{FOPI05}. 
}
\label{fig11}
\end{figure}
\begin{figure}[thb]
\includegraphics[width=6.5cm]{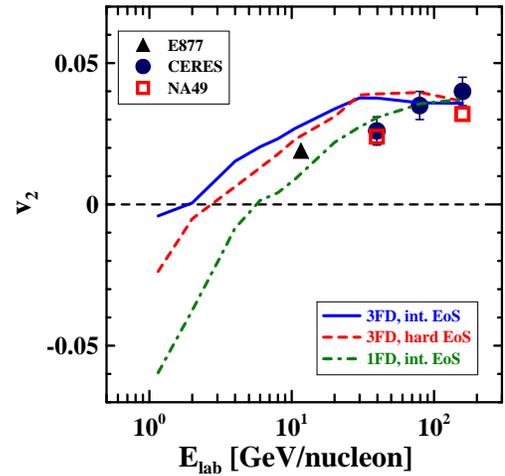}
\caption{(Color online) 
The same as in Fig. \ref{fig11} but for pions. 
}
\label{fig12}
\end{figure}

Results of 1FD simulations are also presented in Fig. \ref{fig8}. 
In fact, at energies $E_{\scr{lab}}=2A,4A$ and 6$A$ GeV the degree of
stopping is rather high already in the 3FD. As it is seen from
Fig. \ref{fig9}, proton rapidity distributions in 1FD do not differ
very much from those in 3FD and from experimental data 
\cite{E895:piKp,E917:piKp}. The same concerns other observables
except flow. The flow reveals much harder pattern within the 1FD at
the same intermediate EoS, cf. also Fig. \ref{fig6}, and effectively
looks like that with the hard EoS in 3FD.

Available data on the pion elliptic flow at the AGS energies, cf.
Fig. \ref{fig10}, are not very spectacular. They allow only to
estimate the order of magnitude of the $v_2$ value. As seen, here the
difference between hard and intermediate EoS's is not already that
large, and the 1FD result is in the best agreement with the data.

Rapidity dependence of the  $v_2$ flow for protons and pions at 
$E_{\scr{lab}}=$ 40$A$ and 158$A$ GeV are presented in Fig. 
\ref{fig7a}. Here we see that stiffness of the EoS only weakly
affects the result. 
Even the difference between extremes, i.e. the
hard EoS and gas EoS, is quite moderate. 
The same feature was observed in the JAM
simulations \cite{IOSN05}, but the reason there was different. In
Ref. \cite{IOSN05}, string excitations dominate particle production at
early times. These strings are not affected by the nuclear mean field
by the assumption of the JAM model. Therefore, the mean-field effects
are small at the SPS energies. In our case the mean-field effects 
(i.e. stiffness of the EoS) are
present to the full extent, and nevertheless their effect is small.

The  pion elliptic flow at 158$A$ GeV 
is in surprisingly good agreement with experimental 
data both in magnitude and shape. However, 
in spite of the weak dependence on the EoS stiffness, we still
essentially overestimate the proton $v_2$ at both incident energies,
40$A$ and 158$A$ GeV. Moreover, the better we reproduce $v_1$ (with
extrasoft EoS), the
worse result we obtain for $v_2$. This again implies that
the {\it nonequilibrium} transverse-momentum anisotropy at the initial 
stage of nuclear collision (now already without serious interplay with
momentum-dependent forces) is very important here. 
An additional argument in favor of this conclusion is that 
cascade simulations \cite{Fuchs,IOSN05,BBSG99} reproduce
the proton $v_2$ quite successfully at 158$A$ GeV, although still
overestimating at 40$A$ GeV. 

The case of energy of 40$A$ GeV deserves special attention. All the
calculations done up to now \cite{IOSN05,SBBSZ04}, in JAM, UrQMD and
HSD, do not reproduce the ``collapse'' of $v_2$ at 40$A$ GeV,
which can be interpreted as the onset of first-order phase transition
\cite{SBBSZ04}. As for the 3FD model, the
disagreement with the data is the largest at 40$A$ GeV,
thus also indicating that something special happens in the system at
this energy.

The earlier pion $v_2$ data of the NA49 collaboration, taken at smaller
acceptance at the top SPS energy \cite{NA49-98-v1}, have been already 
analyzed within hydrodynamic approaches.    
In the expansion model \cite{KSH99} the Bjorken scaling
solution \cite{B83} was assumed for longitudinal evolution and 2D hydro was
solved numerically for transverse one. In this way, the elliptic flow
could be estimated only at the midrapidity point. 
The full 3D expansion
model with postulated initial conditions was applied to the meson elliptic
flow by Hirano \cite{H01}. In a qualitative agreement with
Ref. \cite{KSH99}, it was found that $\rho$-meson decays result in 
almost vanishing azimuthal
anisotropy of pions near the midrapidity. Note that decays of all
relevant resonances are taken into account in our model.

Figs. \ref{fig11} and  \ref{fig12} summarize our results on
$v_2$. 
Besides, they show that $v_2$ becomes low sensitive not only to
the stiffness of the EoS but also to the nuclear stopping power at high
incident energies. Indeed, the 1FD results become very similar to
that of the 3FD model\footnote{Note however that the
1FD results are completely unreasonable for other observables at high
incident energies.}, especially for protons. However, all these
results overestimate the proton $v_2$ at high
incident energies. 
As seen, the 3FD with hard EoS and 1FD properly reproduce the
change of sign of proton $v_2$. 
Unfortunately,
data on pion $v_2$ are practically absent at AGS energies.

\section{Discussion and Conclusions}

In this paper we presented the analysis the collective transverse
flow of nuclear matter within the 3FD model \cite{3FD,3FDm}. This is
not a dedicated study of flow, aiming to reproduce experimental data
by fitting model parameters. On the contrary, we have used model parameters
of our previous paper \cite{3FD}, 
which were fixed to reproduce a great 
body of experimental data in the incident energy range
$E_{\scr{lab}}\simeq$ (1--160)$A$ GeV.
We have varied these parameters only in order to study sensitivity of
the flow to stiffness of the EoS and to the nuclear
stopping power. 

This paper, as well as the previous one \cite{3FD}, is based on a
simple purely hadronic EoS \cite{gasEOS}. 
We do not expect that this EoS perfectly reproduces all the data (and,
in fact, it does not). However, we believe that some useful conclusions
can be drawn from success or failures of simulations based on this
simple EoS.

At the AGS energies the flow turns out to be sensitive to the stopping
power of nuclear 
matter rather than only to the stiffness of the EoS. The stronger
stopping power is, the harder dynamics is, i.e. it looks as if a
harder EoS is employed. This was demonstrated by comparing the 3FD and
1FD calculations.

When the stopping power is fixed to reproduce other observables, 
the flow data favor
more and more soft EoS with the incident energy rise. 
This EoS softening occurs
in the AGS energy range, at higher (SPS) energies the same extrasoft
EoS remains preferable. 
This observation indirectly agrees with that obtained in microscopic
RQMD~\cite{LPX99}, UrQMD~\cite{BBSG99},  
JAM \cite{IOSN05} and HSD \cite{SBBSZ04} models. 
In these models an effective softening of the EoS occurs 
at the early stage of the collision, 
because of gradual (governed by form-factors) 
\cite{SBBSZ04} or abrupt \cite{BBSG99} switching off mean-field
interactions, or
since mean fields turn out to be essentially reduced 
because of string formation \cite{LPX99,IOSN05}. 
With the incident energy rise, 
string excitations and mean-field switching off 
result in progressing softening of the EoS. 
With this mechanism of effective softening these models efficiently
reproduce the directed flow in the whole incident-energy range with
the same mean-field interaction. 
Since the mean-field switching off is accompanied by dominating string
excitations also in models \cite{BBSG99,SBBSZ04}, it is possible to
summarize this mechanism as ``a transition from hadronic to string
matter'', following Ref. \cite{Sahu00}.
Moreover, as it was stated in Ref. \cite{Sahu00}, 
the ``transition from hadronic to string matter'' practically
saturates at the top AGS energy. 
This also is in agreement with our observation that no extra softening
of the EoS in required at SPS energies, as compared with the top AGS
energy.

\begin{figure*}[ht]
\includegraphics[width=11.cm]{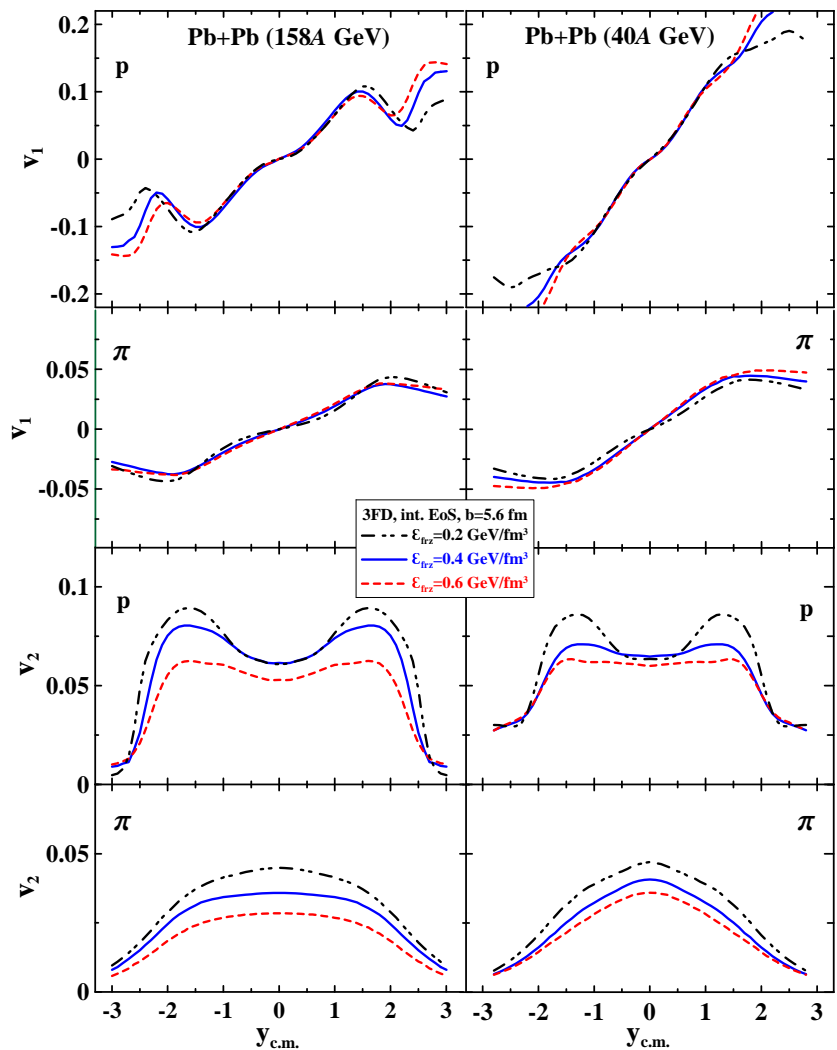}
\caption{(Color online) 
Directed (four upper panels) and elliptic (four lower panels) 
flow of protons  and charged pions in 
mid-central Pb+Pb collisions at $E_{\scr{lab}}=$ 158$A$ GeV (left panels) and 
40$A$ GeV (right panels)  as a function of rapidity. 3FD calculations 
with intermediate EoS 
at $b=$ 5.6 fm are presented for three different freeze-out criteria: 
the freeze-out energy density 
$\varepsilon_{\scr{frz}} =$ 0.2 GeV/fm$^3$ (late freeze-out), 
$\varepsilon_{\scr{frz}} =$ 0.4 GeV/fm$^3$ (our conventional freeze-out), 
and $\varepsilon_{\scr{frz}} =$ 0.6 GeV/fm$^3$ (early freeze-out). }
\label{fig15}
\end{figure*}

We have also found that it is impossible to simultaneously
reproduce the directed  and elliptic flow with the same EoS. 
The directed flow requires a softer EoS,
while the elliptic flow demands for a harder EoS. 
This observation is similar to that done in
Refs. \cite{dani98,DLL02}. In Refs. \cite{Sahu02,IOSN05}
it was found that a proper momentum dependence 
in the nuclear mean field is of prime importance for the 
simultaneous reproduction of the directed and elliptic flows. 
The fact that this momentum dependence is vitally important apparently
suggests that the initial stage of nuclear collision (i.e. the
formation of so-called ``initial fireball'') is essentially
nonequilibrium and that the flow is sensitive to this early
nonequilibrium in transverse-momentum distribution.

At SPS energies the directed-flow data favor the same
extrasoft EoS which was the best beginning from the incident
energy of 8$A$ GeV. This EoS softening can be associated with either
the ``transition from hadronic to string matter'' \cite{Sahu00}, which
has been mentioned already, or could be a signal of deconfinement
transition. 
Within the 3FD simulations 
the elliptic flow becomes low sensitive to the stiffness of the EoS
and even to the stopping power at these energies. 
Lack of the proper description of the {\it nonequilibrium}
transverse-momentum anisotropy at the initial 
stage of nuclear collision prevents the 3FD from good reproduction of
the proton elliptic flow at SPS energies. 
Disagreement of the 3FD with the data is largest at 40$A$ GeV incident
energy, 
which may indicate that something special happens in the system at
this energy.

The above discussion suggests that the flow is an "early-stage" 
observable, i.e. that determined by early-stage evolution of the collision. 
Our calculations with different freeze-out criteria 
(different freeze-out energy densities $\varepsilon_{\scr{frz}}$), displayed 
in Fig. \ref{fig15}, confirm this conjecture. From the point of view of 
freeze-out, the difference between 
$\varepsilon_{\scr{frz}}=$ 0.2 and 0.6 GeV/fm$^3$ is huge ---
they correspond to completely different patterns of the freeze-out. 
Nevertheless, the directed flow turns out to be fairly 
insensitive to this  difference. Slight sensitivity is observed only 
in spectator regions. The elliptic flow is a more subtle quantity. 
It is a measure of difference of the in-plane and out-of-plane flow. 
Taking into account that our calculated $v_2$ are of the order of  
5\% and that sensitivity of $v_1$ with respect to above 
$\varepsilon_{\scr{frz}}$ variations is $\approx$ 1\%, the value of 
20\% for the variation of $v_2$ under above change of 
$\varepsilon_{\scr{frz}}$ looks quite natural. 
In fact, this low sensitivity of the flow to the freeze-out stage is not 
surprising. The $v_1$ flow is a measure of {\it collective momentum} 
accumulated by matter during the expansion stage. The driving force 
of this collective momentum is the pressure gradient created at the early 
compression stage of the collision. If the freeze-out occurs not 
too early (e.g., not right after the compression stage), this 
pressure gradient has 
enough time to accelerate the matter, and hence late-stage evolution 
does not noticeably change the earlier-accumulated collective momentum. 
 
Thus, problems in reproduction of flow data apparently result from 
not quite accurate 
description of  early-stage  transverse-momentum nonequilibrium 
within the 3FD rather than  
are caused by our intermediate hadronic EoS. 
Therefore, keeping in mind that 
with our intermediate hadronic EoS we have succeeded to reasonably
reproduce a large body of experimental data in the incident energy
range $E_{\scr{lab}}\simeq$ (1--160)$A$ GeV \cite{3FD},   
we can still conclude that this EoS is certainly good (however, not
perfect) in this energy range. 
However, since the required EoS softening can be still associated with
a deconfinement transition,  other, more
sophisticated EoS's, including phase transition to the
quark-gluon phase, should be tested within 3FD simulations. 
The present calculations provide a natural benchmark for the future
analysis.

\vspace*{5mm} {\bf Acknowledgements} \vspace*{5mm}

We are grateful to M.~Gazdzicki,
V.V.~Skokov, V.D. Toneev,  and D.N. Voskresensky for fruitful
discussions. 
This work was supported in part by the Deutsche  
Forschungsgemeinschaft (DFG project 436 RUS 113/558/0-3), the
Russian Foundation for Basic Research (RFBR grant 06-02-04001 NNIO\_a),
Russian Federal Agency for Science and Innovations 
(grant NSh-8756.2006.2).

\appendix

\section{Hadronic EoS}
\label{Hadronic EoS}

The EoS, used in present simulations, was originally proposed in
Ref. \cite{gasEOS}. The energy density and pressure are constructed as
follows: 
\begin{eqnarray}
\label{E} 
\hspace*{-5mm}
\varepsilon (n_B,T)&=&
\varepsilon_{\scr{gas}}(n_B,T)+W(n_B),
\\
\label{P}
\hspace*{-5mm}
P(n_B,T)&=&P_{\scr{gas}}(n_B,T)+n_B\frac{dW(n_B)}{dn_B}-W(n_B),
\end{eqnarray}
where $\varepsilon_{\scr{gas}}(n_B,T)$ and $P_{\scr{gas}}(n_B,T)$
are the energy density and pressure of relativistic hadronic gas,
respectively, which depend on baryon density $n_B$ and temperature
$T$. The only difference from the ideal gas is that baryons are
affected by a mean field $U(n_B)=dW(n_B)/dn_B$, 
i.e. the energy of the $a$-baryon
of mass $M_a$ with momentum ${\bf p}$ is  
$\epsilon_a=({\bf p}^2+M_a^2)^{1/2}+b_a U(n_B)$, where $b_a$ is the
baryon number of the $a$-particle. The potential energy density
$W(n_B)$ is parametrized as follows
\begin{eqnarray}
\label{U(n)} 
\hspace*{-11mm}
&&W(n_B)+\varepsilon_{\scr{gas}}(n_B,0)
\cr
\hspace*{-11mm}
&&= m_{N}n_0\left[
a\left( \frac{n_B}{n_{0}}\right)^{5/3}
-b\left(\frac{n_B}{n_{0}}\right)^{2}
+ c \left( \frac{n_B}{n_{0}}\right) ^{7/3}\right]
, 
\end{eqnarray}
where $\varepsilon_{\scr{gas}}(n_B,0)$ is the Fermi energy density of
the cold nucleon gas\footnote{This Fermi energy density is present
  here because in fact we parametrized the zero-temperature part of
  the energy density, $\varepsilon(n_B,0)$.}.
This potential energy density depends only on the density $n_B$. 
Parameters $a$, $b$, and $c$ are determined from the
condition that the cold nuclear matter saturates at $n_0=$ 0.15
fm$^{-3}$ and $\varepsilon (n_0,T=0)/n_0 - m_N=-16$ MeV, and
incompressibility of this nuclear matter is $K$. 
Basically we use the value $K=$ 210 MeV (intermediate EoS). 
However, some
calculations are also done for $K=$ 380 MeV (hard EoS), 
$K=$ 130 MeV (soft EoS) and $K=$ 100 MeV (extrasoft EoS).

Parametrization (\ref{U(n)}) results in superluminal sound velocity
at high baryon densities. To preserve causality at high $n_B$, the
following form of the energy density
\begin{eqnarray}
\label{Eca}
\varepsilon (n_B,T\!=\!0)= n_{0}m_{N}\left[ A\left(
\frac{n_B}{n_{0}}\right)^{2}+C+B\left( \frac{n_{0}}{n_B}\right)
\right]
\end{eqnarray}
is used at $n_B/n_0 >x_c$. Parameters $A$, $B$ and $C$ are
determined on the condition that $\varepsilon (n_B,T=0)$ and its
two first derivatives are continuous at $n_B/n_0 = x_c$.
The baryon density, where parametrization (\ref{Eca}) starts,
depends on the incompressibility $K$: $x_c=6$ for our basic intermediate EoS,
$x_c=5$ for the hard EoS, $x_c=12$ for the soft EoS. Parametrization
(\ref{U(n)}) for the extrasoft EoS does not violate causality at 
considered densities.

\end{document}